\documentclass[3p,,preprint,12pt]{elsarticle}
\makeatletter\if@twocolumn\PassOptionsToPackage{switch}{lineno}\else\fi\makeatother

\usepackage{tabulary,xcolor}
\usepackage{amsfonts,amsmath,amssymb}
\usepackage[T1]{fontenc}
\makeatletter
\let\save@ps@pprintTitle\ps@pprintTitle
\def\ps@pprintTitle{\save@ps@pprintTitle\gdef\@oddfoot{\footnotesize\itshape \null\hfill\today}}
\def\hlinewd#1{%
  \noalign{\ifnum0=`}\fi\hrule \@height #1%
  \futurelet\reserved@a\@xhline}

\AtBeginDocument{\ifNAT@numbers \biboptions{sort&compress}\fi}
\makeatother

\usepackage{ifluatex}
\ifluatex
\usepackage{fontspec}
\defaultfontfeatures{Ligatures=TeX}
\usepackage[]{unicode-math}
\unimathsetup{math-style=TeX}
\else 
\usepackage[utf8]{inputenc}
\fi 
\ifluatex\else\usepackage{stmaryrd}\fi

\usepackage{url,multirow,morefloats,floatflt,cancel,tfrupee}
\makeatletter

\AtBeginDocument{\@ifpackageloaded{textcomp}{}{\usepackage{textcomp}}}
\makeatother
\usepackage{colortbl}
\usepackage{xcolor}
\usepackage{pifont}
\usepackage[nointegrals]{wasysym}
\makeatletter

\def\mcWidth#1{\csname TY@F#1\endcsname+\tabcolsep}

\def\cAlignHack{\rightskip\@flushglue\leftskip\@flushglue\parindent\z@\parfillskip\z@skip}
\def\rAlignHack{\rightskip\z@skip\leftskip\@flushglue \parindent\z@\parfillskip\z@skip}

\usepackage{ifxetex}
\ifxetex\else\if@twocolumn\usepackage{dblfloatfix}\fi\fi

\AtBeginDocument{
\expandafter\ifx\csname eqalign\endcsname\relax
\def\eqalign#1{\null\vcenter{\def\\{\cr}\openup\jot\m@th
  \ialign{\strut$\displaystyle{##}$\hfil&$\displaystyle{{}##}$\hfil
      \crcr#1\crcr}}\,}
\fi
}

\AtBeginDocument{%
  \@ifpackageloaded{endfloat}%
   {\renewcommand\efloat@iwrite[1]{\immediate\expandafter\protected@write\csname efloat@post#1\endcsname{}}}{\newif\ifefloat@tables}%
}%

\def\BreakURLText#1{\@tfor\brk@tempa:=#1\do{\brk@tempa\hskip0pt}}
\let\lt=<
\let\gt=>
\def\processVert{\ifmmode|\else\textbar\fi}

\@ifundefined{subparagraph}{
\def\subparagraph{\@startsection{paragraph}{5}{2\parindent}{0ex plus 0.1ex minus 0.1ex}%
{0ex}{\normalfont\small\itshape}}%
}{}

\newcommand\role[1]{\unskip}
\newcommand\aucollab[1]{\unskip}
  
\@ifundefined{tsGraphicsScaleX}{\gdef\tsGraphicsScaleX{1}}{}
\@ifundefined{tsGraphicsScaleY}{\gdef\tsGraphicsScaleY{.9}}{}
\def\checkGraphicsWidth{\ifdim\Gin@nat@width>\linewidth
	\tsGraphicsScaleX\linewidth\else\Gin@nat@width\fi}

\def\checkGraphicsHeight{\ifdim\Gin@nat@height>.9\textheight
	\tsGraphicsScaleY\textheight\else\Gin@nat@height\fi}

\def\fixFloatSize#1{}
\let\ts@includegraphics\includegraphics

\def\inlinegraphic[#1]#2{{\edef\@tempa{#1}\edef\baseline@shift{\ifx\@tempa\@empty0\else#1\fi}\edef\tempZ{\the\numexpr(\numexpr(\baseline@shift*\f@size/100))}\protect\raisebox{\tempZ pt}{\ts@includegraphics{#2}}}}

\AtBeginDocument{\def\includegraphics{\@ifnextchar[{\ts@includegraphics}{\ts@includegraphics[width=\checkGraphicsWidth,height=\checkGraphicsHeight,keepaspectratio]}}}

\DeclareMathAlphabet{\mathpzc}{OT1}{pzc}{m}{it}

\def\URL#1#2{\@ifundefined{href}{#2}{\href{#1}{#2}}}

\def\UrlOrds{\do\*\do\-\do\~\do\'\do\"\do\-}%
\g@addto@macro{\UrlBreaks}{\UrlOrds}

\edef\fntEncoding{\f@encoding}

\makeatother

\usepackage{float}

\begin{document}

\begin{frontmatter}
	
\title{Python script used as simulator for the teaching of electric field in electromagnetism course
}
    
\author[affadd8b0923bbf33684b1c7a7da8177f5d]{K L Cristiano}
\ead{karen.cristiano@saber.uis.edu.co}
\author[affadd8b0923bbf33684b1c7a7da8177f5d]{D A Triana\corref{contrib-ac1aaf8d6693f7581c60849cc0305c61}}
\ead{dantrica@saber.uis.edu.co}\cortext[contrib-ac1aaf8d6693f7581c60849cc0305c61]{Corresponding author.}
\author[affadd8b0923bbf33684b1c7a7da8177f5d]{A. Estupi{\~{n}}\'{a}n}
\ead{alex.estupinan@saber.uis.edu.co}
    
\address[affadd8b0923bbf33684b1c7a7da8177f5d]{Escuela de F\'{\i}sica\unskip, 
    Laboratorio de Electromagnetismo\unskip, Universidad Industrial de Santander\unskip, Calle 9 cra 27\unskip, Bucaramanga\unskip, Santander\unskip, Colombia}

\begin{abstract}
We present this work like software tool developed in Python, based on a methodology to obtain the electric field produced by n charges. The tool was developed and implemented in courses of electromagnetism and laboratory in three institutions of higher education. The aim for this work is to incorporate information and communication technologies (ICTs) at the university, in accordance with the programs promoted by the Colombian Ministry of Education. We wanted to connect the students with sensitives experiences of the physical phenomena that allow them to improve their experience of learning of subjects traditionally studied through the board course.
\end{abstract}
\end{frontmatter}
    
\section{Introduction}
The authors of the more popular books of electromagnetism like Serway, Sears, Giancoli, and Burbano work the electric field theme from Faraday's view. They study the electric field as a disturbance represented by lines or tubes that propagate through space interacting with the electric charge and finally at a point in space they are pretty well described with a vector \unskip~\cite{339938:7521746}. It's these point where headache of the students begins since they have to combine the vector analysis with the electric field physics description. Usually, both abstract and imaginary concepts can't be bound with one previous sense experience which the students have had.

The other hand, the Colombian Ministry of Education sponsor the use of information and communication technologies (ICTs) as a didactic tool for teaching and learning\unskip~\cite{339938:7522276}. Some pedagogical strategies like Just in Time Teaching (JiTT) take into consideration the use of (ICTs) such as Google classroom, blackboard, and Moodle \unskip~\cite{339938:7522275}. Another initiative from the University of Colorado founded by Nobel Laureate Carl Wieman\unskip~\cite{339938:7522528}. They develop interactive simulations such a way that the students learn by exploring. So the aim to connect the students with sense experiences to improve him learn\unskip~\cite{339938:7522530}. So, arose the idea to develop a methodology that allowed made follow up on the electric field problems at the same time it's viewed from software tool.
    
\section{Methods}
The board class allowed generating discussion with the students, about what should be the appropriate way to solve the problems of the electric field. Then, an algorithm or strategy to provide the solution on 7 steps was defined, independently from electric charges position on the space:

\begin{itemize}
  \item \relax Define the system of reference. It's more intuitive for students to catch on the rectangular coordinates.
  \item \relax Draw the position vectors pointed both the electric charges and the position vector to the point where the electric field will be calculated. These vectors come from the origin of the coordinate system.
  \item \relax Draw the relative position vectors $\vec{r}_i $, that correspond to n differences between the vector at the point and vector at the electric charge.
\end{itemize}
  
\let\saveeqnno\theequation
\let\savefrac\frac
\def\dispfrac{\displaystyle\savefrac}
\begin{eqnarray}
\let\frac\dispfrac
\gdef\theequation{1}
\let\theHequation\theequation
\label{disp-formula-group-a655322cd9cdfe794dac3081943c6159}
\begin{array}{@{}l}\vec{r}_i=\vec{r}_p-\vec{r}_{qi}\end{array}
\end{eqnarray}
\global\let\theequation\saveeqnno
\addtocounter{equation}{-1}\ignorespaces

\begin{itemize}
  \item \relax Find n magnitudes of the n relative position vectors.
\end{itemize}
  
\let\saveeqnno\theequation
\let\savefrac\frac
\def\dispfrac{\displaystyle\savefrac}
\begin{eqnarray}
\let\frac\dispfrac
\gdef\theequation{2}
\let\theHequation\theequation
\label{disp-formula-group-4f750b15b106a4c84367aad62068362c}
\begin{array}{@{}l}r_i = \sqrt{\vec{r}_i \cdot \vec{r}_i}\end{array}
\end{eqnarray}
\global\let\theequation\saveeqnno
\addtocounter{equation}{-1}\ignorespaces

\begin{itemize}
  \item \relax Find the unit vectors of the n relative position vectors.
\end{itemize}
  
\let\saveeqnno\theequation
\let\savefrac\frac
\def\dispfrac{\displaystyle\savefrac}
\begin{eqnarray}
\let\frac\dispfrac
\gdef\theequation{3}
\let\theHequation\theequation
\label{disp-formula-group-88515215847f92e1dab7b06fa926bcda}
\begin{array}{@{}l}\hat{r}_i = \frac{\vec{r}_i}{r_i}\end{array}
\end{eqnarray}
\global\let\theequation\saveeqnno
\addtocounter{equation}{-1}\ignorespaces

\begin{itemize}
  \item \relax From Coulomb law calculate n electric fields. At the practice, it will never distinguish each one, we always will see the total electric field at the point $(x_p, y_p, z_p) $.
\end{itemize}
  
\let\saveeqnno\theequation
\let\savefrac\frac
\def\dispfrac{\displaystyle\savefrac}
\begin{eqnarray}
\let\frac\dispfrac
\gdef\theequation{4}
\let\theHequation\theequation
\label{disp-formula-group-07b035ff78947d1259dad3233749931c}
\begin{array}{@{}l}\vec{E}_i = \frac{q_i}{4\pi \epsilon_0 r_i^{2}}\hat{r}_i\end{array}
\end{eqnarray}
\global\let\theequation\saveeqnno
\addtocounter{equation}{-1}\ignorespaces

\begin{itemize}
  \item \relax From superposition principle, find it the one electric field vector at the point and its magnitude.
\end{itemize}
  
\let\saveeqnno\theequation
\let\savefrac\frac
\def\dispfrac{\displaystyle\savefrac}
\begin{eqnarray}
\let\frac\dispfrac
\gdef\theequation{5}
\let\theHequation\theequation
\label{disp-formula-group-80d98bdde650e2ef380a9c357586cb56}
\begin{array}{@{}l}\vec{E}_p= \sum_i{\vec{E}_i}\end{array}
\end{eqnarray}
\global\let\theequation\saveeqnno
\addtocounter{equation}{-1}\ignorespaces 
Then, we choose Python to write a method named \textit{field\_punctual\_charge} that would include the previous steps. Python is a free language of high level, oriented to objects, focused on the code readability of easy learning, was the ideal language for the construction of the interactive tool.
    
\section{Results}
The method follows the algorithm structure implemented with a while loop that responds to the number of electric charges. On the other hand, the loads, the position vectors with respect to the origin and the relative position vectors are represented in a graph constructed from the module \textit{matplotlib.pyplot}.

\bgroup
\fixFloatSize{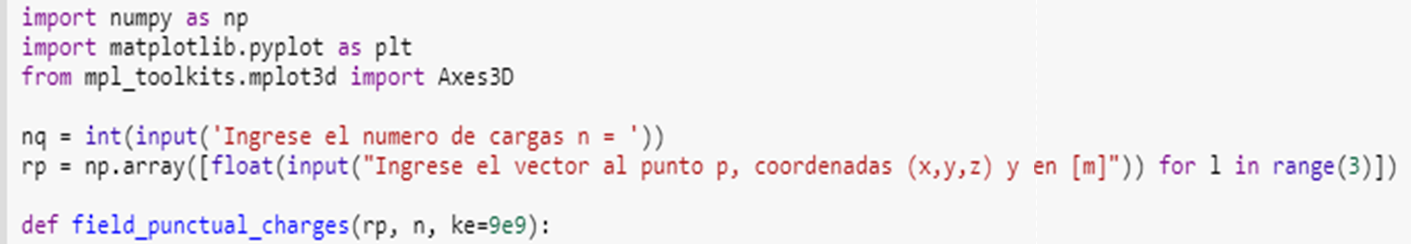}
\begin{figure*}[!htbp]
\centering \makeatletter\IfFileExists{36e36e01-ca55-4b9b-a02c-5aac30980988-upy.png}{\includegraphics{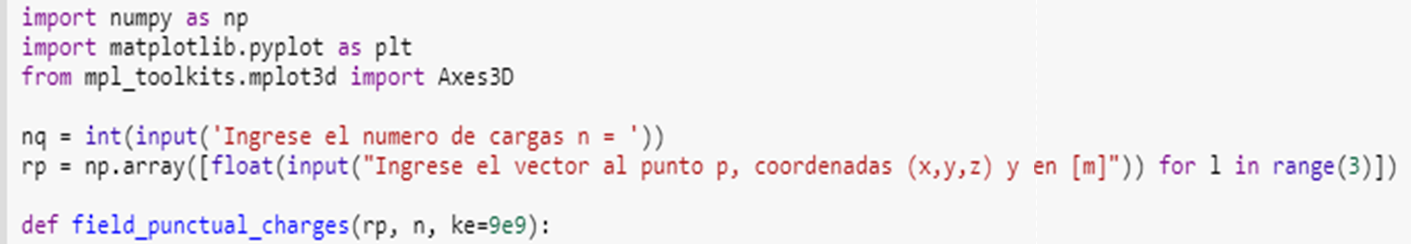}}{}
\makeatother 
\caption{{Python Method that includ the methodology proposed.}}
\label{figure-5cb35f4a881951f01ae20818f81272f0}
\end{figure*}
\egroup
The Python script allows solving problems of the more several charges such as a cubic crystalline structure formed by positive ions (see figure 3a) indeed straight line charge and others too. An advantage over the Phet simulator is the ability to represent and solve three-dimensional charge configurations (see figures 3b and 3c).

\bgroup
\fixFloatSize{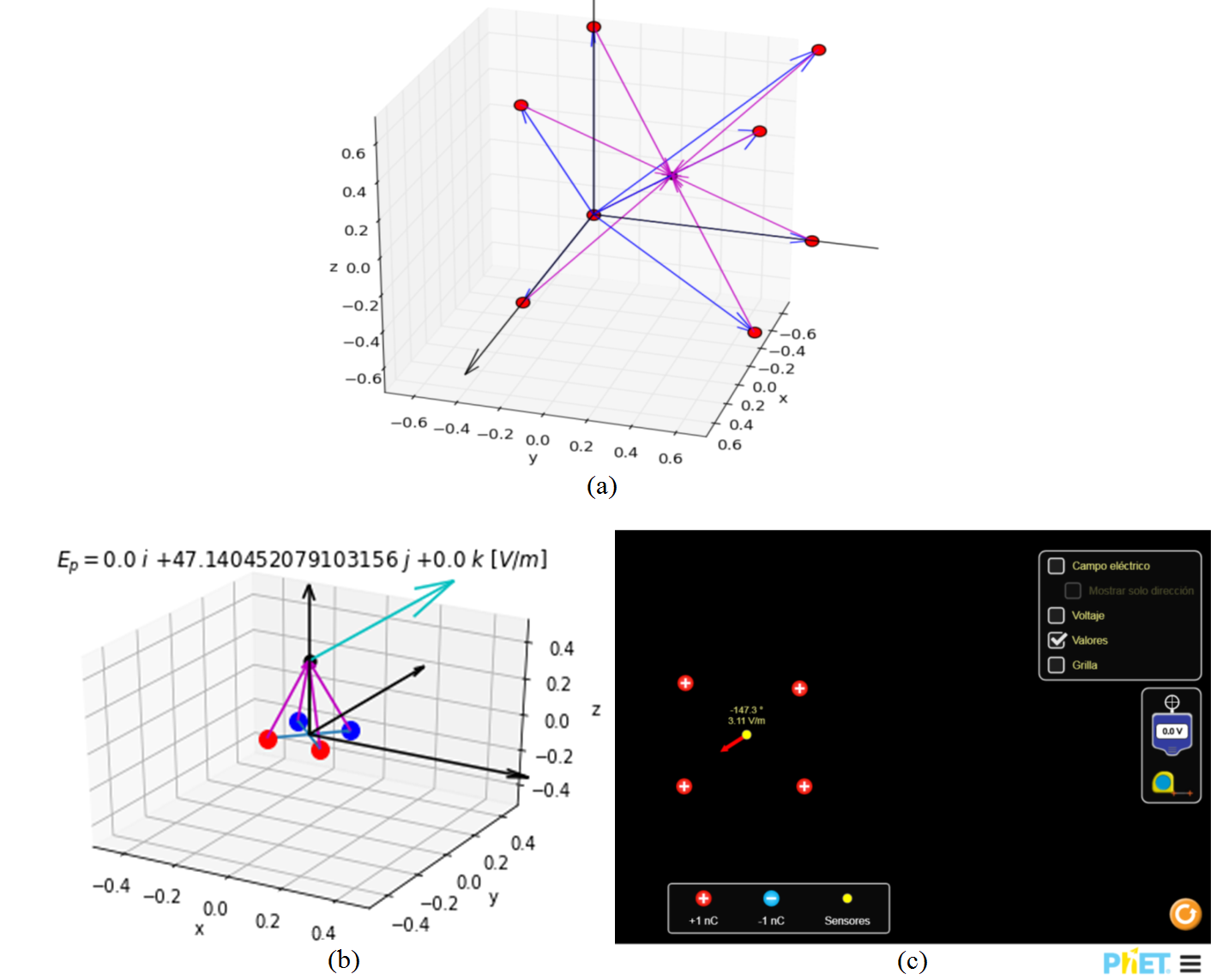}
\begin{figure*}[!htbp]
\centering \makeatletter\IfFileExists{e22c1eae-9f44-43b8-9f6a-e7726c3d5189-ugraf.png}{\includegraphics{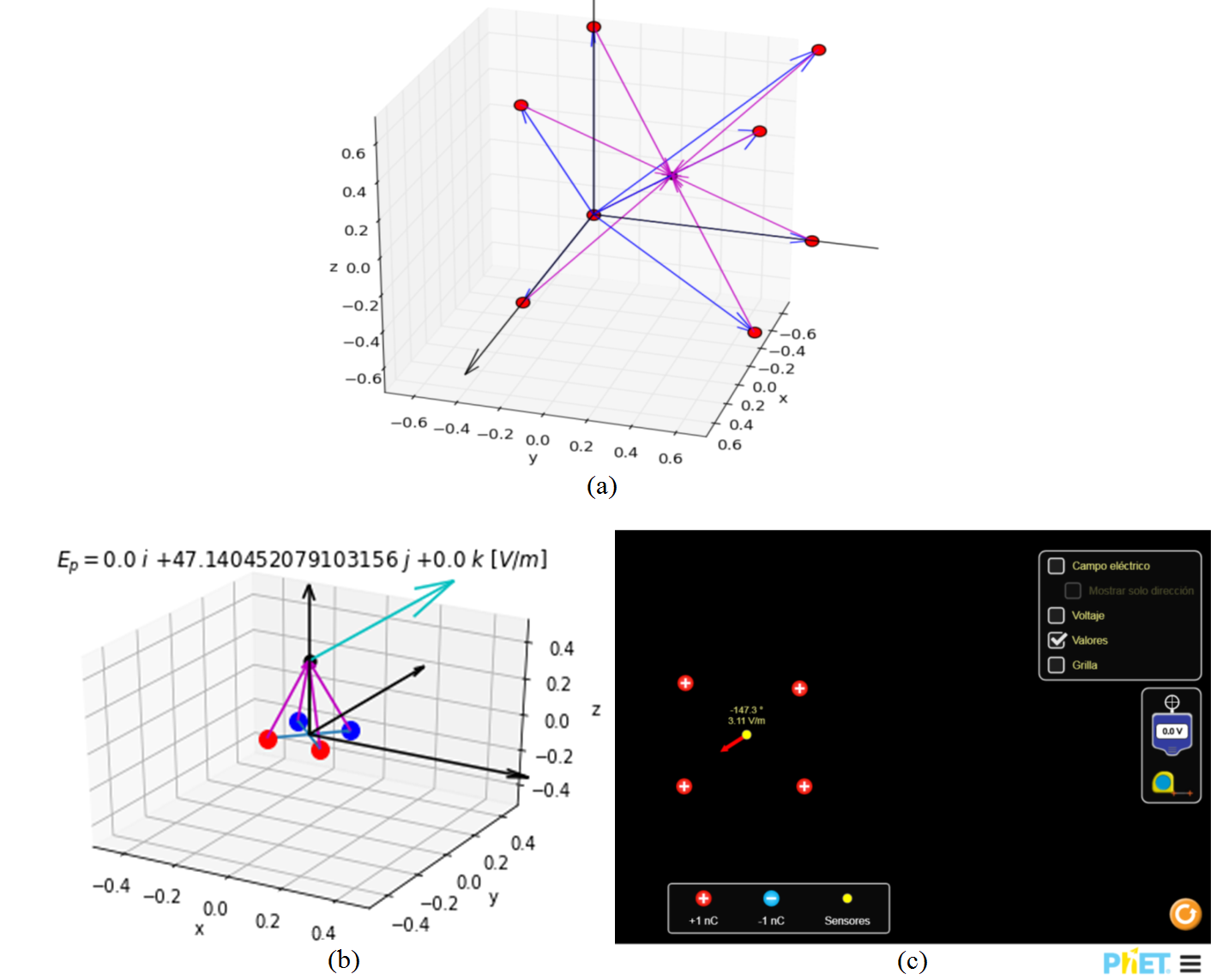}}{}
\makeatother 
\caption{{Cubic (a) and square (b) configurations of electric charge, both carried out in own simulator. (c) Square configuration of electric charges made in Phet simulator.}}
\label{figure-ae4795f13162cf52ac10cb8824058447}
\end{figure*}
\egroup
Positive electric charges were drawn in red while negative ones in blue, The point p where the electric field was calculated it has been drawn in black, the position vectors in blue and relative position vectors in magenta.

At the same time what the plot is obtained, it's possible to review in the terminal the magnitudes of the electric charges, the position vectors of the electric charges, the relative position vectors and their magnitude and unit vector, to finish with the electric field vectors produced for each electric charge and the electric field vector from the superposition.
    
\newpage    
    
\section{Conclusions}
A methodology and method were developed in Python to calculate the electric field produced by n electric charges, with which it's also possible to calculate electric fields produced by continuous load distributions.

The methodology and the Python script were implemented into electromagnetism class (theoretical and laboratory) in three Universities. It improved the experience of students accustomed to traditional teaching methods in their previous courses of Calculus and Mechanics. The code is available on GitHub at the following link: \url{https://github.com/alexestupinan123/Teaching_physics_Python.git}.

\newpage

\section*{Acknowledgements}We offer special thanks to our research group (CIMBIOS), to the School of Physics from UIS for motivating the use of (ICTs) in the subjects, thanks too UNAB and UTS, for being creative laboratories.


\end{document}